\documentclass[twocolumn,aps,prd,epsf,showpacs]{revtex4-1}
\pdfoutput=1
\usepackage[pdftex]{graphicx}
\usepackage{color}
\usepackage{colortbl}
\usepackage{xcolor}
\usepackage{amsmath}
\usepackage{newtxtext,newtxmath}
\usepackage{amsfonts}
\usepackage{dsfont}
\usepackage[]{units}
\usepackage{braket}
\usepackage[utf8]{inputenc}
\usepackage[T1]{fontenc}
\usepackage{url}
\usepackage[english]{babel}
\usepackage{bm}
\usepackage{url}
\usepackage{verbatim}
\usepackage[]{cleveref}
\usepackage{bm}
\usepackage{slashed}
\usepackage{array,multirow}
\usepackage{lipsum}
\usepackage{algorithm}
\usepackage[noend]{algpseudocode}
\algnewcommand\algorithmicforeach{\textbf{for each}}
\algdef{S}[FOR]{ForEach}[1]{\algorithmicforeach\ #1\ \algorithmicdo}

\newcommand\Tstrut{\rule{0pt}{2.9ex}}         
\newcommand\Bstrut{\rule[-1.2ex]{0pt}{0pt}}   
\newcommand\TBstrut{\Tstrut\Bstrut}           

\newcommand{\be}{\begin{equation}}
\newcommand{\ee}{\end{equation}}
\newcommand{\bea}{\begin{eqnarray}}
\newcommand{\eea}{\end{eqnarray}}

\newcommand{\bei}{\begin{itemize}}

\newcommand{\eei}{\end{itemize}}
\newcommand{\ben}{\begin{enumerate}}
\newcommand{\een}{\end{enumerate}}

\newcommand{\gtapprox}{\raisebox{-0.5ex}{$\,\stackrel{>}{\scriptstyle\sim}\,$}}

\graphicspath{{figures/}}

\begin{document}

\title{ 
Computation of the quarkonium and meson-meson composition of the $\Upsilon(nS)$ states \\ and of the new $\Upsilon(10753)$ Belle resonance from lattice QCD static potentials
}

\author{$^{(1)}$Pedro Bicudo}
\email{bicudo@tecnico.ulisboa.pt}

\author{$^{(1)}$Nuno Cardoso}
\email{nuno.cardoso@tecnico.ulisboa.pt}

\author{$^{(2)}$Lasse Müller}
\email{lmueller@itp.uni-frankfurt.de}

\author{$^{(2),(3)}$Marc Wagner}
\email{mwagner@itp.uni-frankfurt.de}

\affiliation{\vspace{0.1cm}$^{(1)}$CeFEMA, Dep.\ F\'{\i}sica, Instituto Superior T\'ecnico, Universidade de Lisboa, Av.\ Rovisco Pais, 1049-001 Lisboa, Portugal}

\affiliation{\vspace{0.1cm}$^{(2)}$Goethe-Universit\"at Frankfurt, Institut f\"ur Theoretische Physik, Max-von-Laue-Stra{\ss}e 1, D-60438 Frankfurt am Main, Germany}

\affiliation{\vspace{0.1cm}$^{(3)}$Helmholtz Research Academy Hesse for FAIR, Campus Riedberg, Max-von-Laue-Stra{\ss}e 12, D-60438 Frankfurt am Main, Germany}

\begin{abstract}
We compute the composition of the bottomonium $\Upsilon(nS)$ states (including $\Upsilon(10860)$) and the new $\Upsilon(10753)$ resonance reported by Belle in terms of quarkonium and meson-meson components. We use a recently developed novel approach utilizing lattice QCD string breaking potentials for the study of resonances. This approach is based on the diabatic extension of the Born Oppenheimer approximation and the unitary emergent wave method and allows to compute the poles of the $\mbox{S}$ matrix. We focus on $I=0$ bottomonium $S$ wave bound states and resonances, where the Schr\"odinger equation is a set of coupled differential equations. One of the channels corresponds to a confined heavy quark-antiquark pair $\bar b b$, the others to pairs of heavy-light mesons. In a previous study only one meson-meson channel $\bar{B}^{(\ast)} B^{(\ast)}$ was considered. Now we also include the closed strangeness channel $\bar{B}_s^{(\ast)} B_s^{(\ast)}$ extending our formalism significantly to have a more realistic description of bottomonium. We confirm the new Belle resonance $\Upsilon(10753)$ as a dynamical meson-meson resonance with around $76 \%$ meson-meson content. Moreover, we identify $\Upsilon(4S)$ and $\Upsilon(10860)$ as states with both sizable quarkonium and meson-meson contents. With these results we contribute to the clarification of ongoing controversies in the vector bottomonium spectrum.
\end{abstract}


\maketitle


\section{\label{sec:intro}Introduction}

Starting from the determination of lattice QCD static potentials with dynamical quarks, our long term goal is a complete computation of masses and decay widths of bottomonium bound states and resonances as poles of the $\mbox{S}$ matrix. We expect our technique to be eventually updated to study the full set of exotic $X$, $Y$ and $Z$ mesons. In this work, however, we focus on the somewhat simpler, but nevertheless controversial $I = 0$ bottomonium $S$ wave resonances.

It has been standard for many years to use the ordinary static potential obtained in lattice QCD as the confining quarkonium potential to study the heavy quarkonium spectrum \cite{Godfrey:1985xj}. Recently we developed a novel approach to apply lattice QCD string breaking potentials (as e.g.\ computed in Ref.\ \cite{Bali:2005fu}) to coupled channel systems, opening the way for the computation of the spectrum and the composition of resonances with heavy quarks \cite{Bicudo:2019ymo}. This approach allows to study hadronic interactions non-perturbatively with input from first principles lattice QCD.

In the past, microscopic determinations of hadronic strong interactions, i.e.\ from quark interactions, were mostly addressed with quark models.
From the onset of QCD, while developing the quark bag model, Jaffe predicted multiquarks such as tetraquarks. Moreover, he started computing microscopically potentials for coupled channels of hadrons \cite{Jaffe:1976ig,Jaffe:1976ih}. An important type of potentials are the hadron-hadron potentials, such as $V_{\bar M M}$ we use in this work, where the number of quarks is preserved. The microscopic computation of hadron-hadron potentials in models needs to include the different algebraic color, flavor, spin and three dimensional space or momentum factors, the latter being systematically performed with the resonating group method \cite{Ribeiro:1981fk}.
A second type of potentials are the mixing potentials, such as the $V_\text{mix}$ used in this work, which couples channels with a different number of quarks and microscopically includes the creation or annihilation of a light quark-antiquark pair. In quark models it was realized a long time ago from the symmetries the QCD vacuum that the pair creation has ${^S L_J} = {^3 P_0}$ quantum numbers in spectroscopic notation \cite{Micu:1968mk,LeYaouanc:1972vsx}. 
Spontaneous chiral symmetry breaking can also be included in the  ${^3 P_0}$ pair creation mechanism \cite{Bicudo:1989sj}.
This microscopic knowledge contributed to very successful models with quark-antiquark channels and meson-meson channels allowing to study a large number of resonances \cite{Kokoski:1985is}, complicated dynamical resonances \cite{vanBeveren:1986ea} and resonances with heavy quarks \cite{Bruschini:2020voj}. 
These models share a vanishing interaction between two mesons, but have different mixing potentials. For instance Ref. \cite{vanBeveren:1986ea} uses a delta-shell potential for simplicity and Ref. \cite{Bruschini:2020voj} uses a Gaussian shell potential with has an additional parameter. Refs.\ \cite{Kokoski:1985is,Bicudo:1989sj} compute microscopically $V_\text{mix}$ from the overlap of the meson wave functions and the ${^3 P_0}$ pair creation term. 

However, since multiquarks may be very complex systems, we expect some of them to be sensitive to the details of the potentials. Unfortunately, the potentials are not fully fixed by the symmetries of QCD, i.e.\ what one can infer is only qualitative. For example $V_\text{mix}$ at large quark-antiquark separations must decay rapidly like the meson wave functions in the overlap, exponentially or as a Gaussian. At small separations, due to the momentum or position present in the ${^3 P_0}$ mechanism, it must linearly increase from $0$. Moreover, due to the parity in the ${^3 P_0}$ mechanism, the orbital angular momenta of the quark-antiquark channel and the meson-meson channel must differ by $1$. Obviously, these constraints do not fully determine the potentials and there are certain degrees of freedom left. Lattice QCD, on the other hand, is a method to unambiguously compute these potentials from QCD. Thus, we expect that the computation of the potentials $V_{\bar Q Q}$ (the quark-antiquark potential), $V_\text{mix}$ and $V_{\bar M M}$ with lattice QCD will allow to clarify important aspects of certain quarkonium resonances and experimentally observed multiquarks.

Lattice QCD computations fully incorporate the dynamics of the light quarks and gluons, while the heavy quarks are approximated as static color charges. The dynamics of the heavy quarks is then added in a second step using techniques from quantum mechanics as in the Born-Oppenheimer approximation \cite{Born:1927}. Due to heavy quark symmetry the spin of the heavy quarks is conserved \cite{Isgur:1988gb,Isgur:1989ed,Isgur:1991wq,Georgi:1990um}.
In previous works this Born-Oppenheimer approach was applied to investigate exotic mesons containing a bottom and an anti-bottom quark. For example, the spectrum of $\bar{b} b$ hybrid mesons was studied extensively (see e.g.\ Refs.\ \cite{Juge:1999ie,Braaten:2014qka,Berwein:2015vca,Capitani:2018rox}) using static potentials computed within pure SU(3) lattice gauge theory, which are confining and do not allow decays to pairs of lighter mesons. The first application of the Born-Oppenheimer approach using meson-meson potentials computed in full lattice QCD to study tetraquarks can be found in Refs.\ \cite{Bicudo:2012qt,Brown:2012tm}. For instance the existence of a stable $\bar b \bar b u d$ tetraquark with quantum numbers $I(J^P) = 0(1^+)$ was confirmed \cite{Bicudo:2015kna,Bicudo:2016ooe}, whereas other flavor combinations do not seem to form four-quark bound states \cite{Bicudo:2015vta}. In this context the approach was also updated by including techniques from scattering theory and a $\bar b \bar b u d$ tetraquark resonance with quantum numbers $I(J^P) = 0(1^-)$ was predicted \cite{Bicudo:2017szl}.
The Born-Oppenheimer approach approach should also allow for the inclusion of the heavy quark spin, either from the experimental hyperfine splitting \cite{Bicudo:2016ooe} or with lattice QCD computations of $1 / m_b$ corrections to the static potentials \cite{Koma:2006fw}. 

In this work we continue within this framework and significantly extend our recent study of systems with a heavy quark and a heavy antiquark and possibly another light quark-antiquark pair \cite{Bicudo:2019ymo}. This constitutes an even more challenging system, which might open the way to study bottonomium $X$, $Y$ and $Z$  resonances. The approach then requires the lattice QCD determination of several potentials including a $\bar b b$ potential, a $\bar{B}^{(*)} B^{(*)}$ potential and a mixing potential, and allows the study of resonances and their decays with scattering theory.
In this work we do not carry out such lattice QCD computations, but use results from an existing study of string breaking \cite{Bali:2005fu}.
Notice that we go beyond the Born-Oppenheimer adiabatic approximation \cite{Braaten:2014qka}, since the quarkonium potential crosses open meson-meson thresholds. Formally, our system is then denominated diabatic \cite{PhysRev.131.229,PhysRev.179.111,Bruschini:2020voj}, since the heavy quarks are much slower that the light degrees of freedom, but the state of the heavy quarks nevertheless changes, when a decay occurs.

The main goal of this paper is to compute the composition of $I = 0$ bottonomium resonances in terms of quarkonium $\bar{Q} Q$, and meson-meson $\bar{M} M$ components. It is important to note that in Ref.\ \cite{Bicudo:2019ymo} only a $\bar{B}^{(\ast)} B^{(\ast)}$ meson-meson channel was included. However, since the closed strangeness $\bar{B}^{(\ast)} B^{(\ast)}$ channel is very close to the bottomomium resonances we are interested in, we also include this channel in the present work.
Clearly this case is more involved, because there are three coupled channels, a confined quarkonium channel with flavor $\bar{b} b$ and the two meson-meson decay channels with flavor $\bar b b ( \bar u u + \bar d d)/\sqrt 2$ and $\bar b b \bar s s$.

In Table~\ref{tab:bottomonium} we show the available experimental results according to the Review of Particle Physics \cite{Zyla:2020zbs}. Since we work in the heavy quark limit, the heavy quark spins $S_Q^{PC}$ do not appear in the Hamiltonian and the relevant quantum numbers $\widetilde J^{PC}$ are the remaining part of the total angular momentum and the corresponding parity and charge conjugation (also listed in Table~\ref{tab:bottomonium}). Notice that we also list several states observed at Belle with large significance \cite{Mizuk:2012pb,Abdesselam:2019gth}. These states are not yet confirmed by other experiments, because presently Belle and Belle~II are the only experiments designed to study bottomonium.

\begin{table}[t!]
\begin{ruledtabular}
\begin{tabular}{cc|lc|c}
name  & $I^G (J^{PC})$ & $m$ [MeV] &  $\Gamma$  [MeV] & $ \widetilde{J}^{P C}$ \\ 
 \hline
$\eta_b (1S)$ & $0^+(0^{-+})$ & \hspace{2pt} $9399.0 \pm 2.3$ & $10 \pm 5$ & $0^{++}$\\ 
$\Upsilon (1S)$ & $0^-(1^{--})$ & \hspace{2pt} $9460.30 \pm 0.26$ & $(54.02 \pm 1.25)10^{-3}$ & $0^{++}$\\ 
$\chi_{b0} (1P)$ & $0^+(0^{++})$ & \hspace{2pt} $9859.44 \pm 0.73$ & - & $1^{--}$\\ 
$\chi_{b1} (1P)$ & $0^+(1^{++})$ & \hspace{2pt} $9892.78 \pm 0.57$ & - & $1^{--}$\\ 
$h_{b} (1P)$ & $?^?(1^{+-})$ & \hspace{2pt} $9899.3 \pm 0.8$ & -  & $1^{--}$\\ 
$\chi_{b2} (1P)$ & $0^+(2^{++})$ & \hspace{2pt} $9912.21 \pm 0.57$ & -  & $1^{--}$\\ 
$\eta_b(2S)_\text{\tiny Belle}$ & $0^+(0^{-+})$ & \hspace{2pt} $9999.0\pm 6.3$  & - & $0^{++}$\\
$\Upsilon (2S)$ & $0^-(1^{--})$ &  $10023.26 \pm 0.31$ & $(31.98\pm 2.63)10^{-3}$  & $0^{++}$\\ 
$\Upsilon (1D)$ & $0^-(2^{--})$ &  $10163.7 \pm 1.4$ & -  & $2^{++}$\\ 
$\chi_{b0} (2P)$ & $0^+(0^{++})$ &  $10232.5 \pm 0.9$ & -  & $1^{--}$\\ 
$\chi_{b1} (2P)$ & $0^+(1^{++})$ &  $10255.46 \pm 0.77$ & -  & $1^{--}$\\ 
$h_{b} (2P)_\text{\tiny Belle}$ & $?^?(1^{+-})$ & $10259.8 \pm 1.6$ & -  & $1^{--}$\\ 
$\chi_{b2} (2P)$ & $0^+(1^{++})$ &  $10268.65 \pm 0.72$ & -  & $1^{--}$\\ 
$\Upsilon (3S)$ & $0^-(1^{--})$ &  $10355.2 \pm 0.5$ & $(20.32\pm 1.85)10^{-3}$  & $0^{++}$\\ 
$\chi_{b1} (3P)$ & $0^+(1^{++})$ &  $10512.1 \pm 2.3$ & -  & $1^{--}$\\ 
\hline
$\Upsilon(4S)$ & $0^-(1^{--})$ &  $10579.4 \pm 1.2$ & $20.5\pm 2.5$  & $0^{++}$\\
\hline 
$\Upsilon(10753)_\text{\tiny Belle}$ & $0^-(1^{--})$ &  $10752.7 \pm 7.0$ & \hspace{4pt}$35.5\pm 21.6$  & $0^{++}$\\ 

$\Upsilon(10860)$ & $0^-(1^{--})$ &  $10885.2 \pm2.1$ & $37\pm 4$  & $0^{++}$\\ 
$\Upsilon(11020)$ & $0^-(1^{--})$ &  $10000\pm 4 $ & $24\pm 7$  & $0^{++}$\\ 
\end{tabular}
\end{ruledtabular}
\caption{\label{tab:bottomonium}Masses $m$ and decay widths $\Gamma$ of $I = 0$ bottomonium according to the Review of Particle Physics \cite{Zyla:2020zbs}. We also include several states observed at Belle \cite{Mizuk:2012pb,Abdesselam:2019gth}, but not yet confirmed by other experiments. We add an extra column with the quantum numbers $\widetilde{J}^{PC}$ conserved in the infinite quark mass limit (in the last three lines $\widetilde{J}^{PC} = 2^{++}$ is also a possibility). We mark with horizontal lines the opening of the $\bar B B$ and $\bar B^* B^*$ thresholds.}
\end{table}

In particular a new resonance, $\Upsilon(10753)$, possibly another $\Upsilon(nS)$ state or a $Y$ state, since it is a vector but suggested to be of exotic nature, has recently been observed at Belle with a mass around $10.75 \, \text{GeV}$ \cite{Abdesselam:2019gth}. The previously observed resonances $\Upsilon(4S)$ and $\Upsilon(10860)$ approximately match quark model predictions of bottomonium and, thus, this new resonance comes in excess and needs to be understood.

Notice also that the discovery of this resonance by Belle with the process $e^+ e^- \to \Upsilon (nS) \pi^+\pi^-$ resulted from the experimental effort to clarify the controversy on the nature of the other excited $\Upsilon$ resonances \cite{Abdesselam:2019gth}. The $\Upsilon(4S)$, $\Upsilon(10860)$, and $\Upsilon(11020)$, although having masses approximately compatible with the quark model, have transitions to lower bottomonia with the emission of light hadrons with much higher rates compared to expectations for ordinary bottomonium. A possible interpretation is that these excited $\Upsilon$ states have large admixtures of $\bar{B}^{(*)} B^{(*)}$ meson pairs \cite{Meng:2007tk,Simonov:2008ci,Voloshin:2012dk,Sungu:2018iew}. Another scenario is that they do not correspond to the $S$ wave states $\Upsilon(5S)$ and $\Upsilon(6S)$, but instead to the $D$ wave states $\Upsilon(3D)$ and $\Upsilon(4D)$ \cite{Li:2019qsg,Liang:2019geg,Giron:2020qpb}. The Belle experiment was, thus, designed to produce and study $\Upsilon$ states with a large $\bar{B}^{(*)} B^{(*)}$ admixture.

After the observation of the new resonance at Belle, more exotic interpretations have been proposed for the excited $\Upsilon$ states. Most interpretations consider the new $\Upsilon(10753)$ resonance as a non-conventional state, e.g.\ a tetraquark \cite{Wang:2019veq,Ali:2019okl} or a hybrid meson \cite{TarrusCastella:2019lyq,Chen:2019uzm,Brambilla:2019esw}. There are, however, also different interpretations, e.g.\ in Ref.\ \cite{Liang:2019geg} it is claimed that the $\Upsilon(4S)$ is not a simple quarkonium state. 

In this work, we aim to contribute to the clarification of the controversies concerning the bottomonium resonances $\Upsilon(4S)$, $\Upsilon(10753)$ and $\Upsilon(10860)$. While the low-lying bottomonium spectrum up to the $\bar{B} B$ threshold was studied within full lattice QCD extensively \cite{Meinel:2009rd,Meinel:2010pv,Dowdall:2011wh,Aoki:2012xaa,Lewis:2012ir,Dowdall:2013jqa,Wurtz:2015mqa,1810330}, it is extremely difficult to investigate higher resonances in a similar setup, in particular those with several decay channels. Thus, as already explained above, we continue our recent work \cite{Bicudo:2019ymo} using lattice QCD potentials and applying the emergent wave method to study $I=0$ bottomonium $S$ wave resonances. Using this strategy, independently of the experimental observation of the resonance $\Upsilon(10753)$ at Belle \cite{Abdesselam:2019gth}, which we were not aware of at that time, we predicted a similar resonance with mass $10774^{+4}_{-4} \, \text{MeV}$ \cite{Bicudo:2019ymo}.
We now extend this work including another important meson-meson channel, the $\bar{B}^{(*)}_s B^{(*)}_s $ channel, with threshold between the $\Upsilon(10753)$ and $\Upsilon(10860)$. Within this improved setup we determine the composition of all bound states and resonances up to $\approx 11 \, \text{GeV}$, i.e.\ the percentage of a pair of confined heavy quarks $\bar{b} b$ as well as the percentage of a pair of heavy-light mesons $\bar{B}^{(*)} B^{(*)}$ and $\bar{B}^{(*)}_s B^{(*)}_s$.

This paper is structured as follows. In Section~\ref{sec:approach} we review the theoretical basics of our approach from Ref.\ \cite{Bicudo:2019ymo}. We discuss, how to utilize lattice QCD static potentials, and how to solve the coupled Schr\"odinger equation to obtain a quarkonium and one or two meson-meson wave functions. We also review our results for the poles of the $\mbox{S}$ matrix, i.e.\ for $I=0$ bottomonium $S$ wave resonances. In Section~\ref{sec:content} we propose a technique to determine the percentage of the quark-antiquark and the meson-meson component of a bottomonium state, either a bound state (if we neglect heavy quark annihilation and electroweak interactions) or a resonance. Then we apply this technique to $\Upsilon(1S)$, $\Upsilon(2S)$, $\Upsilon(3S)$, $\Upsilon(4S)$, $\Upsilon(10753)$ and $\Upsilon(10860)$. In Section~\ref{sec:content} we also discuss results within the two channel setup, i.e.\ considering quarkonium $\bar b b$ and a meson pair $\bar{B}^{(*)} B^{(*)}$, and in Section~\ref{sec:content3} we discuss results within the three channel setup, i.e.\ with an extra $\bar{B}_s^{(*)} B_s^{(*)}$ channel. In Section~\ref{sec:conclu} we conclude.


\section{Summary of our approach \label{sec:approach}}

In this section we briefly summarize our approach from Ref.\ \cite{Bicudo:2019ymo} to study quarkonium resonances with isospin $I = 0$ in the diabatic extension of the Born-Oppenheimer approximation, using lattice QCD static potentials. We  also recapitulate the main results from Ref.\ \cite{Bicudo:2019ymo}. Moreover, we extend the approach to three coupled channels, including a $\bar{B}^{(*)}_s B^{(*)}_s$ channel.


\subsection{Theoretical basics -- two coupled channels}

We consider systems composed of a heavy quark-antiquark pair $\bar{Q} Q$ and either no light quarks (quarkonium) or another light quark-antiquark pair $\bar{q} q$ with isospin $I = 0$ (for large $\bar{Q} Q$ separation two heavy-light mesons $M = \bar{Q} q$ and $\bar{M} = \bar{q} Q$). We treat the heavy quark spins as conserved quantities such that the energy levels of $\bar{Q} Q (\bar{q} q)$ systems as well as their decays and and resonance parameters do not depend on these spins. Moreover, we assume that two of the four components of the Dirac spinors of the heavy quarks $Q$ and $\bar{Q}$ vanish. These approximations become exact for static quarks and are expected to yield reasonably accurate results for $b$ quarks, possibly even for $c$ quarks.

In Ref.\ \cite{Bicudo:2019ymo} we have derived in detail a coupled channel Schr\"odiger equation for a 4-component wave function $\psi(\mathbf{r}) = (\psi_{\bar{Q} Q}(\mathbf{r}) , \vec{\psi}_{\bar{M} M}(\mathbf{r}))$ (Eq.\ (10) in Ref.\ \cite{Bicudo:2019ymo}). The upper component of this wave function represents the $\bar{Q} Q$ channel, the lower three components represent the $\bar{M} M$ channel. For the $\bar{M} M$ channel we consider only the lightest heavy-light mesons with $J^P = 0^-$ and $J^P = 1^-$, i.e.\ $B$ and $B^\ast$ mesons for $Q \equiv b$ (as usual, $J$, $P$ and $C$ denote total angular momentum, parity and charge conjugation). Within the approximations stated above these two mesons have the same mass. One can show that the spin of the two light quarks is $1$, which is represented by the three components of $\vec{\psi}_{\bar{M} M}(\mathbf{r})$. Note that we ignore decays of $\bar{Q} Q$ to lighter quarkonium and a light $I = 0$ meson, e.g.\ a $\sigma$ or an $\eta$ meson, because they are suppressed by the OZI rule.

$\widetilde{J}^{P C}$ denotes total angular momentum excluding the heavy quark spins and the corresponding parity and charge conjugation. It is a conserved quantity. As in Ref.\ \cite{Bicudo:2019ymo} we focus throughout this work on $\widetilde{J}^{P C} = 0^{+ +}$. Thus $J^{P C} = S_Q^{P C}$, where $S_Q$ denotes the heavy quark spin, with only two possibilities, $S_Q^{P C} = 0^{- +} , 1^{- -}$.

\begin{figure}[htb]
\includegraphics[width=0.49\textwidth]{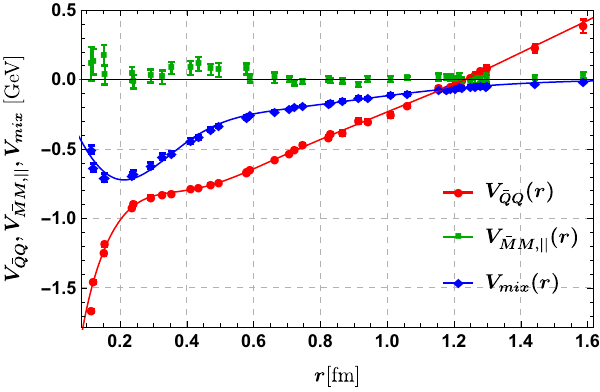}
\caption{ (Color online.) Potentials $V_{\bar{Q} Q}(r)$, $V_{\bar{M} M,\parallel}(r)$ and $V_{\textrm{mix}}(r)$ as functions of the $\bar{Q} Q$ separation $r$. The curves correspond to the parameterizations (\ref{eq:fitVQQ}) to (\ref{eq:fitMix}) with parameters listed in Table~\ref{tab:fitsGevFm}.
\label{fig:fit_V}}
\end{figure}

The coupled channel Schr\"odinger equation for the partial wave with $\widetilde{J} = 0$ is a 2-channel equation,
\newpage
\begin{widetext}
\begin{eqnarray}
\nonumber & & \left(-\frac{1}{2} \left(\begin{array}{cc} 1/\mu_Q & 0 \\ 0 & 1/\mu_M \end{array}\right) \partial_r^2 + \frac{1}{2 r^2} \left(\begin{array}{cc} 0 & 0 \\ 0 & 2/\mu_M \end{array}\right) + V_0(r) + 2 m_M - E\right)
\left(\begin{array}{c} u(r) \\ \chi_{\bar M M}(r) \end{array}\right) =
-\left(\begin{array}{c} V_\textrm{mix}(r) \\ V_{\bar{M} M,\parallel}(r) \end{array}\right) k r j_1(k r) \quad , \\
\label{EQN001_} & & \hspace{0.7cm} V_0(r) = \left(\begin{array}{cc} V_{\bar{Q} Q}(r) & V_\textrm{mix}(r) \\ V_\textrm{mix}(r) & V_{\bar{M} M,\parallel}(r) \end{array}\right) .
\end{eqnarray}
\end{widetext}
The upper equation represents the $\bar{Q} Q$ channel with orbital angular momentum $L_{\bar{Q} Q} = \widetilde{J} = 0$. $u(r)$ is the radial part of the $\widetilde{J} = 0$ partial wave of the wave function
\begin{eqnarray}
\psi_{\bar{Q} Q}(\mathbf{r}) = \sqrt{4 \pi} i \frac{u(r)}{k r} Y_{0,0}(\Omega) + \ldots
\end{eqnarray}
with the dots $\ldots$ denoting partial waves with $\widetilde{J} > 0$. Similarly, the lower equation represents the $\bar{M} M$ channel with orbital angular momentum $L_{\bar{M} M} = 1$. $j_1(k r)$ and $\chi_{\bar M M}(r)$ are the radial parts of the $\widetilde{J} = 0$ partial waves of the incident plane wave and the emergent spherical wave of the 3-component wave function
\begin{eqnarray}
\vec{\psi}_{\bar{M} M}(\mathbf{r}) = \sqrt{4 \pi} i \bigg(j_1(k r) + \frac{\chi_{\bar M M}(r)}{k r}\bigg) \mathbf{Z}_{\bar M M}(\Omega) + \ldots
\end{eqnarray}
with $\mathbf{Z}_{\bar M M}(\Omega) = \mathbf{e}_r / \sqrt{4 \pi}$ and the dots $\ldots$ denoting partial waves with $\widetilde{J} > 0$. Moreover, $m_Q$ and $m_M$ are the heavy quark and heavy-light meson masses, respectively, and $\mu_Q = m_Q / 2$ and $\mu_M = m_M / 2$ are the corresponding reduced masses. The energy $E$ and the momentum $k$ are related according to $k = \sqrt{2 \mu_M E}$. The potentials $V_{\bar{Q} Q}(r)$, $V_{\bar{M} M,\parallel}(r)$ and $V_\textrm{mix}(r)$ represent the energy of a heavy quark-antiquark pair, the energy of a pair of heavy-light mesons and the mixing between the two channels, respectively. In Ref.\ \cite{Bicudo:2019ymo} we related these potentials algebraically to lattice QCD correlators computed and provided in detail in Ref.\ \cite{Bali:2005fu} in the context of string breaking for lattice spacing $a \approx 0.083 \, \text{fm}$ and pion mass $m_\pi \approx 650 \, \text{MeV}$. The data points for $V_{\bar{Q} Q}(r)$, $V_{\bar{M} M,\parallel}(r)$ and $V_\textrm{mix}(r)$ are shown in Fig.\ \ref{fig:fit_V} together with appropriate parameterizations,
\begin{eqnarray}
\label{eq:fitVQQ} & & V_{\bar{Q} Q}(r) = E_0 - \frac{\alpha}{r} + \sigma r + \sum_{j=1}^2 c_{\bar{Q} Q,j} r \exp\bigg(-\frac{r^2}{2 \lambda_{\bar{Q} Q,j}^2}\bigg) \\
\label{eq:fitVBB} & & V_{\bar{M} M,\parallel}(r) = 0 \\
\label{eq:fitMix} & & V_{\textrm{mix}}(r) = \sum_{j=1}^2 c_{\textrm{mix},j} r \exp\bigg(-\frac{r^2}{2 \lambda_{\textrm{mix},j}^2}\bigg) .
\end{eqnarray}
The parameters appearing in Eq.\ (\ref{eq:fitVQQ}) to Eq.\ (\ref{eq:fitMix}) are collected in Table~\ref{tab:fitsGevFm}.

\begin{table}[htb]
\begin{ruledtabular}
\begin{tabular}{c|c|c}
\TBstrut
potential & parameter & value \\ 
\hline
\Tstrut
$V_{\bar{Q} Q}(r)$ & $E_0$                      & $-1.599(269) \, \textrm{GeV}\phantom{1.^{-1}}$ \\
                   & $\alpha$                   & $+0.320(94) \phantom{1.0 \, \textrm{GeV}^{-1}}$ \\
                   & $\sigma$                   & $+0.253(035) \, \textrm{GeV}^{2\phantom{-}}\phantom{1.}$ \\
                   & $c_{\bar{Q} Q,1}$          & $+0.826(882) \, \textrm{GeV}^{2\phantom{-}}\phantom{1.}$ \\
                   & $\lambda_{\bar{Q} Q,1}$    & $+0.964(47) \, \textrm{GeV}^{-1}\phantom{1.0}$ \\
                   & $c_{\bar{Q} Q,2}$          & $+0.174(1.004) \, \textrm{GeV}^{2\phantom{-}}$ \\
\Bstrut
                   & $\lambda_{\bar{Q} Q,2}$    & $+2.663(425) \, \textrm{GeV}^{-1}\phantom{1.}$ \\
\hline
\TBstrut
$V_{\bar{M} M,\parallel}(r)$ & -- & -- \\
\hline
\Tstrut
$V_{\textrm{mix}}(r)$ & $c_{\textrm{mix},1}$       & $-0.988(32) \, \textrm{GeV}^{2\phantom{-}}\phantom{1.0}$ \\
                      & $\lambda_{\textrm{mix},1}$ & $+0.982(18) \, \textrm{GeV}^{-1}\phantom{1.0}$ \\
                      & $c_{\textrm{mix},2}$       & $-0.142(7) \, \textrm{GeV}^{2\phantom{-}}\phantom{1.00}$ \\
\Bstrut
                      & $\lambda_{\textrm{mix},2}$ & $+2.666(46) \, \textrm{GeV}^{-1}\phantom{1.0}$ \\
\end{tabular}
\end{ruledtabular}
\caption{\label{tab:fitsGevFm}The parameters of the potential parametrizations (\ref{eq:fitVQQ}) to (\ref{eq:fitMix}).}
\end{table}

It is interesting to compare our potentials to those utilized in quark models.
The models of Refs.\ \cite{Kokoski:1985is,vanBeveren:1986ea,Bruschini:2020voj}
all have a confining $V_{\bar{Q} Q}(r)$ and our lattice QCD potential is also confining.
This is not surprising, since confinement is a central feature of most quark models.
However, it is remarkable that the meson-meson interaction $V_{\bar{M} M,\parallel}(r)$ obtained from lattice QCD correlators is compatible with zero within error bars and at the same time all these models have no direct meson-meson interaction as well. In the case of the models this is a simplification, but in our case it is a first principles QCD result. Such a vanishing meson-meson interaction is not universal. It appears in the coupled channel $I=0$ bottomonium system, but for instance not in the $\bar b \bar b u d$ system, where a significant attraction leads to a tetraquark boundstate \cite{Wagner:2010ad,Bicudo:2012qt}. 
In what concerns the mixing potential, the lattice QCD result $V_{\textrm{mix}}(r)$ has a richer structure than those used in Refs.\ \cite{vanBeveren:1986ea,Bruschini:2020voj}, which are non-vanishing only in a certain region of $r$ close to the string breaking distance $r_c$. We note again that our $V_{\textrm{mix}}(r)$ is a first principles QCD result and that there is no physical or phenomenological reason, why this potential should not have the behavior shown in Fig.\ \ref{fig:fit_V}. It vanishes
at large $r$, as in the case of the models, but extends to much smaller quark-antiquark separations than the potentials of Refs.\ \cite{vanBeveren:1986ea,Bruschini:2020voj}. 
Indeed, the lattice QCD result $V_{\textrm{mix}}(r)$ is close to those calculated microscopically with the ${^3 P_0}$ mechanism of Refs.\ \cite{Kokoski:1985is,Bicudo:1989sj}.
Finally we notice that there is a small but clearly visible bump in $V_{\bar{Q} Q}(r)$ at $r \approx 0.25 \, \text{fm}$, which is typically not present in quark model potentials. This bump is a consequence of the non-vanishing mixing between energy eigenstates on the one hand and $\bar{Q} Q$ and $\bar{M} M$ states on the other hand. With lattice QCD the ground state and the first excitation are computed as functions of $r$, where the ground state corresponds to a confining potential without a bump at small $r$ (see e.g.\ Fig.\ 13 in Ref.\ \cite{Bali:2005fu}, the curve labeled ``state $| 1 \rangle$''). Lattice QCD also provides the mixing angle, i.e.\ the contribution of the ground state and the first excitation to the $\bar{Q} Q$ and $\bar{M} M$ states. This mixing moves $V_{\bar{Q} Q}(r)$ and $V_{\bar{M} M,\parallel}(r)$ closer together for non-vanishing mixing angle. The mixing angle is particularly large at separations $r \approx 0.25 \, \text{fm}$ (see Fig.\ 15 in Ref.\ \cite{Bali:2005fu}) as also indicated by the extremum in the mixing potential $V_{\textrm{mix}}(r)$. Thus, the mixing generates a bump in $V_{\bar{Q} Q}(r)$ and removes a similar bump present in the first excitation (see Fig.\ 13 in Ref.\ \cite{Bali:2005fu}, the curve labeled ``state $| 2 \rangle$'') leading to a essentially vanishing meson-meson interaction $V_{\bar{M} M,\parallel}(r)$.

The appropriate boundary conditions for the radial wave functions $u(r)$ and $\chi_{\bar M M}(r)$ are
\begin{eqnarray}
\label{EQNbc1} & & u(r) \propto r \quad \textrm{for } r \rightarrow 0 \\
\label{EQNbc2} & & u(r) = 0 \quad \textrm{for } r \rightarrow \infty \\
\label{EQNbc3} & & \chi_{\bar M M}(r) \propto r^2 \quad \textrm{for } r \rightarrow 0 \\
\label{EQNbc4} & & \chi_{\bar M M}(r) = i t_{\bar M M} k r h_1^{(1)}(k r) \quad \textrm{for } r \rightarrow \infty ,
\end{eqnarray}
where $h_1^{(1)}$ is a spherical Hankel function of the first kind and $t_{\bar M M}$ is the scattering amplitude and an eigenvalue of the $\mbox{S}$ matrix. We compute $t_{\bar M M}$ as a function of the complex energy $E$. Poles of $t_{\bar M M}$ on the real axis below the $\bar{M} M$ threshold indicate bound states. Poles of $t_{\bar M M}$ at energies with non-vanishing negative imaginary parts represent resonances with masses $m = \textrm{Re}(E)$ and decay widths $\Gamma = -2 \textrm{Im}(E)$. $t_{\bar M M}$ is also related to the corresponding scattering phase via $e^{2 i \delta_{\bar M M}} = 1 + 2 i t_{\bar M M}$.


\subsection{Main results from Ref.\ \cite{Bicudo:2019ymo} -- two coupled channels}

In Ref.\ \cite{Bicudo:2019ymo} we applied our approach to study bottomonium bound states and resonances with $I = 0$. For $m_M$, which is the energy reference of our system, we use the spin-averaged mass of the $B$ meson and the $B^*$ meson, i.e.\ $m_M = (m_B + 3 m_{B^*}) / 4 = 5.313 \, \textrm{GeV}$ \cite{Zyla:2020zbs}. $\mu_Q = m_Q/2$ in the kinetic term of the coupled channel Schr\"odinger equation (\ref{EQN001_}) is the reduced mass of the $b$ quark. Since results are only weakly dependent on $m_Q$ (see previous works following a similar approach, e.g.\ Refs.\ \cite{Bicudo:2015kna,Karbstein:2018mzo}), we use for simplicity $m_Q = 4.977 \, \textrm{GeV}$ from quark models \cite{Godfrey:1985xj}.

\begin{figure*}[p]
\begin{center}
two coupled channels: quarkonium and $\bar{B}^{(*)} B^{(*)}$
\end{center}
\includegraphics[width=1.8\columnwidth]{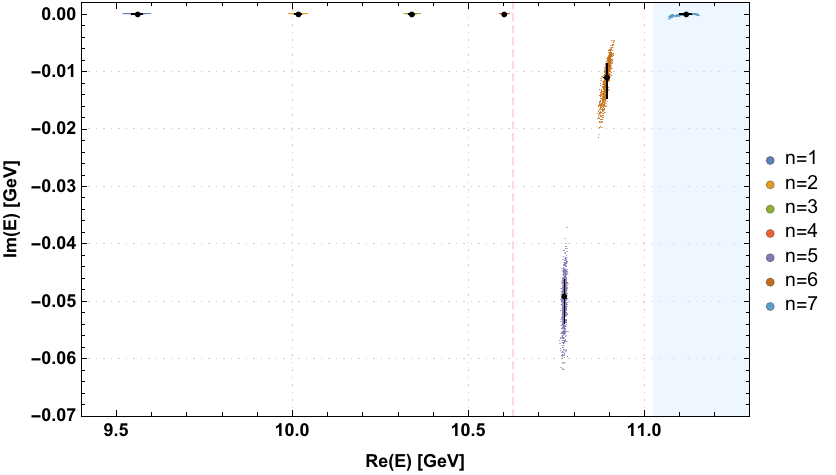} \\

\begin{center}
three coupled channels: quarkonium, $\bar{B}^{(*)} B^{(*)}$ and $\bar{B}_s^{(*)} B_s^{(*)}$
\end{center}
\includegraphics[width=1.8\columnwidth]{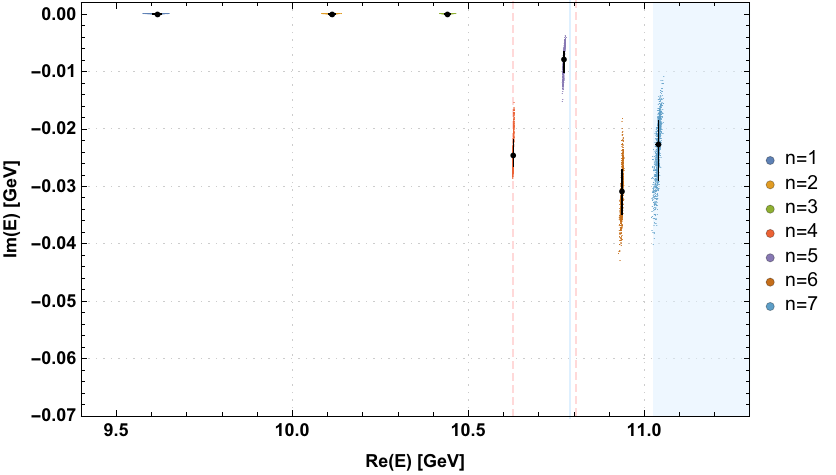}

\caption{(Color online.) 
Positions of the poles in the complex energy plane of $t_{\bar M M}$ for the case of two coupled channels (upper plot) and of the $\mbox{T}$ matrix for the case of three coupled channels (lower plot) for all bound states and resonances below $11.3 \, \textrm{GeV}$. Colored point clouds represent the 1000 resampled sets of lattice QCD correlators, while black points and crosses represent the corresponding mean values and error bars (see Ref.\ \cite{Bicudo:2019ymo} for details). The vertical dashed lines mark the spin-averaged $\bar{B}^{(*)} B^{(*)}$ threshold at $10.627 \, \textrm{GeV}$ and the spin-averaged $\bar{B}_s^{(*)} B_s^{(*)}$ threshold at $10.807 \, \textrm{GeV}$. The shaded region above $11.025 \, \textrm{GeV}$ marks the opening of the threshold of one heavy-light meson with negative parity and another with positive parity, beyond which our results should not be trusted. We also mark with a vertical line $E_\text{threshold} = 10.790 \, \textrm{GeV}$, which corresponds to two times the mass of a static-light meson from Ref.\ \cite{Bali:2005fu}.
\label{fig:errorpole}}
\end{figure*}

\begin{table*}[t!]
\renewcommand{\arraystretch}{1.7}
\begin{ruledtabular}
\begin{tabular}{c|cc|cc||cc|ccc||ccc}
\Tstrut
  & \multicolumn{4}{c||}{from poles of $t_{\bar M M}$, two channels}
  & \multicolumn{5}{c||}{from poles of $\mbox{T}$, three channels}
  & \multicolumn{3}{c}{from experiment} \\
\Bstrut
$n$ & $m$ [GeV] & $\Gamma$ [MeV] & $\% \bar Q Q$ [\%] & $\% \bar M M$ [\%]
   & $m$ [GeV] & $\Gamma$ [MeV] & $\% \bar Q Q$ [\%] & $\% \bar M M$ [\%] & $\% \bar M_s M_s$ [\%]
  & name & $m$ [GeV] & $\Gamma$ [MeV] \\
\hline
\Tstrut
$1$ & & & & & & & & &
  & $\eta_b(1S)$ & $\phantom{0}9.399^{+2}_ {-2}$ & $10^{+5}_ {-4\phantom{0}}$ \\
$1$ &
 $\phantom{0}9.562_{-17}^{+11}$ & 0 & $89_{-0}^{+1}{}_{\phantom{\, -0}}$ & $11_{-1}^{+0}{}_{\phantom{\, -0}}$
 & $\phantom{0}9.618_{-15}^{+10}$ & $0$ &
 $84_{-1\phantom{0}}^{+1}{}_{\phantom{+0}}$ & $12_{-0}^{+0}{}_{\phantom{+0}}$ & $\phantom{0}5_{-0}^{+0}{}_{\phantom{+0}}$
  & $\Upsilon(1S)$ & $\phantom{0}9.460^{+0}_ {-0}$ & $\approx 0$ \\
$2$ &
 $10.018_{-10}^{+8\phantom{0}}$ & 0 & $90_{-0}^{+0}{}_{\phantom{\, -0}}$ & $10_{-0}^{+0}{}_{\phantom{\, -0}}$
  &  $10.114_{-11}^{+7}$ & 0 &  
  $84_{-0\phantom{0}}^{+0}{}_{\phantom{+0}}$ & $12_{-0}^{+0}{}_{\phantom{+0}}$ & $\phantom{0}4_{-0}^{+0}{}_{\phantom{+0}}$
  &  $\Upsilon(2S)$ & $10.023^{+0}_ {-0}$ & $\approx 0$ \\
$3$ &
 $10.340_{-9\phantom{0}}^{+7\phantom{0}}$ & 0 & $88_{-0}^{+0}{}_{\phantom{\, -0}}$ & $12_{-0}^{+0}{}_{\phantom{\, -0}}$
  &  $10.442_{-9\phantom{0}}^{+7}$ & 0 &  
  $79_{-0\phantom{0}}^{+0}{}_{\phantom{+0}}$ & $17_{-0}^{+0}{}_{\phantom{+0}}$ & $\phantom{0}4_{-0}^{+0}{}_{\phantom{+0}}$
  & $\Upsilon(3S)$ & $10.355^{+0}_ {-0}$ & $\approx 0$ \\
%
\cline{6-13} \Tstrut
$4$ &
 $10.603_{-6\phantom{0}}^{+5\phantom{0}}$ & 0 & $70_{-2}^{+3}{}_{\phantom{\, -0}}$ & $30_{-3}^{+2}{}_{\phantom{\, -0}}$
  &  $10.629_{-1\phantom{0}}^{+1}$ & $49.3_{-3.9}^{+5.4}$ &  
  $67_{-5\phantom{0}}^{+0}{}_{-1}^{+1}$ & $29_{-0}^{+5}{}_{-1}^{+1}$ & $\phantom{0}4_{-0}^{+0}{}_{-0}^{+0}$
  & $\Upsilon(4S)$ & $10.579^{+1}_ {-1}$ & $21^{+3\phantom{0}}_ {-3}$ \\
\cline{2-5} \cline{11-13} \Tstrut
\Tstrut
$5$ &
 $10.774_{-4\phantom{0}}^{+4\phantom{0}}$ &
 $98.5_{-5.9}^{+9.2}$ & $\phantom{0}6_{-0}^{+1}{}_{-1}^{+2}$ & $94_{-1}^{+0}{}_{-2}^{+1}$
  &  $10.773_{-2\phantom{0}}^{+1}$ & $15.9_{-4.4}^{+2.9}$ &  
  $24_{-3\phantom{0}}^{+3}{}_{-1}^{+1}$ & $60_{-4}^{+4}{}_{-2}^{+1}$ & $16_{-2}^{+1}{}_{-1}^{+1}$
  & $\Upsilon(10753)$ & $10.753^{+7}_ {-7}$ & $36^{+22}_{-14}$ \\
\cline{2-13} \Tstrut
$6$ &
 $10.895_{-10}^{+7\phantom{0}}$ &
 $22.2_{-4.9}^{+7.1}$ & $59_{-4}^{+4}{}_{-2}^{+2}$ & $41_{-4}^{+4}{}_{-2}^{+2}$
  &  $10.938_{-2\phantom{0}}^{+2}$ & $61.8_{-8.0}^{+7.6}$ &  
  $35_{-7}^{+11}{}_{-3}^{+4}$ & $40_{-6}^{+3}{}_{-3}^{+3}$ & $25_{-6}^{+5}{}_{-0}^{+0}$
  & $\Upsilon(10860)$ & $10.885^{+3}_{-2}$ & $37^{+4\phantom{0}}_{-4}$
\end{tabular}
\end{ruledtabular}
\renewcommand{\arraystretch}{1.0}
\caption{
Masses $m = \textrm{Re}(E_\text{pole})$ and decay widths $\Gamma=-2 \textrm{Im}(E_\text{pole})$ for $I=0$ bottomonium with $\widetilde{J}^{P C} = 0^{+ +}$ from the coupled channel Schr\"odinger equations (\ref{EQN001_}) and (\ref{EQN001}) and the corresponding $\bar Q Q$ and $\bar M M$ or $\bar M_s M_s$ percentages (for $R_\textrm{max} = 2.4 \, \textrm{fm}$). For comparison we also list available experimental results. The $\bar{B}^{(*)} B^{(*)}$ and  $\bar{B}_s^{(*)} B_s^{(*)}$ thresholds are marked by horizontal lines. Errors on our results for $m$ and $\Gamma$ are purely statistical, while for $\% \bar Q Q$, $\% \bar M M$ and $\% \bar M_s M_s$ we additionally show systematic uncertainties for the resonances as discussed in Section~\ref{SEC566}. 
\label{tab:polesemergent}}
\end{table*}

In Ref.\ \cite{Bicudo:2019ymo} we presented both the scattering amplitude $t_{\bar M M}$ and the phase shift $\delta_{\bar M M}$ for real energies $E$ above the $\bar{B}^{(*)} B^{(*)}$ threshold at $10.627 \, \textrm{GeV}$ (throughout this paper we use a notation slightly different from that in Ref.\ \cite{Bicudo:2019ymo}, $t_{\bar M M} \equiv t_{1 \rightarrow 0,0}$ and $\delta_{\bar M M} \equiv \delta_{1 \rightarrow 0,0}$). We also checked probability conservation by showing the Argand diagram for $t_{\bar M M}$. The main numerical results of Ref.\ \cite{Bicudo:2019ymo} are, however, the poles of $t_{\bar M M}$ in the complex energy plane, which are shown in Fig.\ \ref{fig:errorpole} (upper plot) and collected in Table~\ref{tab:polesemergent}.

There are four poles on the real axis below the $\bar{B}^{(*)} B^{(*)}$ threshold representing bound states ($n = 1,\ldots,4$ in Table~\ref{tab:polesemergent}). By comparing them to the experimental results from Table~\ref{tab:bottomonium}, we identify them with $\eta_b(1S) \equiv \Upsilon(1S)$, $\Upsilon(2S)$, $\Upsilon(3S)$ and $\Upsilon(4S)$. We also obtained a resonance around $10.895 \, \textrm{GeV}$, which matches $\Upsilon(10860)$ with experimentally found mass $(10.885 \pm 0.002) \, \text{GeV}$ rather well ($n = 6$ in Table~\ref{tab:polesemergent}). Moreover, in Ref.\ \cite{Bicudo:2019ymo} we predicted a new, dynamically generated resonance close the the $\bar{B}^{(\ast)} B^{(\ast)}$ threshold with mass around $10.774 \, \textrm{GeV}$ ($n = 5$ in Table~\ref{tab:polesemergent}). Recently Belle has observed a bottomonium state at $(10.753 \pm 0.007) \, \textrm{GeV}$ denoted as $\Upsilon(10753)$ not yet confirmed by other experiments, which could correspond to our prediction. 

However, for the $n=5$ and $n=6$ states, which are close in energy to $\Upsilon(10753)$ and $\Upsilon(10860)$, it should be important to also include the $\bar{B}_s^{(*)} B_s^{(*)}$ channel, since its threshold opens between these two states. Thus we proceed by studying three coupled channels and compare the results with those obtained in the two channel case. This will provide insights, how important meson-meson thresholds and the corresponding channels are for resonance properties. We note that to obtain reliable and realistic masses and widths for resonances above $\approx 11.025 \, \textrm{GeV}$, which is the threshold of one heavy-light meson with negative parity and another with positive parity, one has to include even further meson-meson channels.


\subsection{Extension to the three coupled channel case}

We include the $\bar{B}_s^{(*)} B_s^{(*)}$ channel in Eq.\ (\ref{EQN001_}) using the same string breaking potentials as before, i.e.\ those provided by Ref.\ \cite{Bali:2005fu}. We expect this is to be a reasonable approximation, because the mass of the light quarks in Ref.\ \cite{Bali:2005fu} is between the physical $u/d$ and the physical $s$ quark mass. Thus we use the same mixing potential for both channels. 

Moreover, the direct interaction between the static-light meson pairs in Eq.\ (\ref{EQN001_}) turned out to be negligible in the two coupled channel case (see Fig.\ \ref{fig:fit_V} and the detailed discussion in Ref.\ \cite{Bicudo:2019ymo}). Thus we use vanishing meson-meson interactions also in the case of three coupled channels, since one can hardly anti\-cipate a mechanism that increases the meson-meson interaction either between a $B_s^{(*)}$ and a $\bar{B}_s^{(*)}$ or in the transition between a $\bar{B}^{(*)} B^{(*)} $ and a $\bar{B}_s^{(*)} B_s^{(*)}$.

In detail we extend the potential matrix to three coupled channels as follows. From the Review of Particle Physics we get $m_{B_s^0} = 5.367 \, \text{GeV}$ (1 spin state) and $m_{B_s^*} = 5.415 \, \text{GeV}$ (3 spin states). Using spin symmetry, the average is $5.403 \, \text{GeV}$. The $\bar{B}_s^{(*)} B_s^{(*)}$ threshold opens at $10.807 \, \text{GeV}$, indeed between the new $\Upsilon(10753)$ and the $\Upsilon(10860)$.

One can estimate the quark mass used in in the lattice QCD computation of Ref.\ \cite{Bali:2005fu} using a theorem of Partially Conserved Axial Currents (PCAC), applicable to the light quarks $u$, $d$ and $s$. According to the Gell-Mann, Oakes and Renner relation \cite{GellMann:1968rz}
the light current quark masses and the pseudoscalar mesons obey the relation
${m_\text{meson}}^2 {f_\pi}^2 = (m_q + m_{\bar q}) \langle \Omega | \bar q q | \Omega \rangle$ in first order.
We consider the average $u$ and $d$ quark mass $m_l = (m_u + m_d) / 2$ and the charge averaged masses for the pion and the kaon given by $m_\pi = ((2/3) \times 139.6 + (1/3) \times 135.0) \, \text{MeV} = 138.0 \, \text{MeV}$ and $m_\kappa = ((2/4) \times 493.7 + (2/4) \times 497.6) \, \text{MeV} = 495.7 \, \text{MeV}$. We find $m_s = 24.8 \times m_l$. However, the light quark mass used in Ref.\ \cite{Bali:2005fu}, corresponding to the light pseudoscalar meson mass $m_{\pi,\text{Ref.\ \cite{Bali:2005fu}}} = 654.1 \, \text{MeV}$, amounts to $m_{l,\text{Ref.\ \cite{Bali:2005fu}}} = (654.1 / 138.0)^2 m_l = 22.5 \times m_l$. Thus, the light quark mass of Ref.\ \cite{Bali:2005fu} is even closer to the $s$ quark mass than to the physical $u/d$ quark mass. As stated above, we use the mixing potential obtained from the lattice QCD correlators of Ref.\ \cite{Bali:2005fu} for both the $\bar{B}^{(*)} B^{(*)}$ and the $\bar{B}_s^{(*)} B_s^{(*)}$ channel.

Another point is a possibly different algebraic factor for the mixing potential of the new $\bar{B}_s^{(*)} B_s^{(*)}$ channel. We note that the mixing potential is proportional to the lattice QCD creation operator $\mathcal{O}_{\bar M M}^{\Sigma_g^+}$ (see Eqs.\ (14) and (18) in Ref.\ \cite{Bicudo:2019ymo}). For two degenerate flavors $u$ and $d$ this operator is composed of two terms of identical form (one for each flavor), but has to be normalized by another factor $1 / \sqrt{2}$ compared to a single flavor $s$. Thus the mixing potential is weaker by the factor $1 / \sqrt{2}$ for the new $\bar{B}_s^{(*)} B_s^{(*)}$ channel. Alternatively, one can set up a $3 \times 3$ Schr\"odinger equation for the 2-flavor case with a quarkonium, a $\bar{B}_u^{(*)} B_u^{(*)}$ and a $\bar{B}_d^{(*)} B_d^{(*)}$ channel, using a ``1-flavor mixing potential'' $V_\text{mix}^\text{1 flavor}$ to describe the the mixing between the quarkonium and each of the two meson-meson channels. This $3 \times 3$ can be block diagonalized, where a $2 \times 2$ block is identical to Eq.\ (\ref{EQN001_}) and a $1 \times 1$ block corresponds to $I = 1$. The mixing potential appearing the $2 \times 2$ block is $V_\text{mix} = \sqrt{2} V_\text{mix}^\text{1 flavor}$, confirming $V_\text{mix} / \sqrt{2}$ as mixing potential for the new $\bar{B}_s^{(*)} B_s^{(*)}$ channel.

To conclude, with the new $\bar{B}_s^{(*)} B_s^{(*)}$ channel we now have three channels:
$b \bar b$, $\bar{B}^{(*)} B^{(*)}$ and $\bar{B}_s^{(*)} B_s^{(*)}$.
This amounts to adding one more line and column to the Hamiltonian in the coupled-channel Schr\"odinger equation (\ref{EQN001_}), where the threshold in the third component of the wave function is $10.807 \, \text{GeV}$, while the threshold in the second component remains at $10.627 \, \text{GeV}$.
The mixing potential in the new matrix elements $(1,3)$ and $(3,1)$ is weaker by the factor $1 / \sqrt{2}$ compared to the mixing potential $V_\text{mix}$ in the matrix elements $(1,2)$ and $(2,1)$. Moreover, the new matrix elements $(2,3)$ and $(3,2)$ vanish, since there is neither a kinetic energy nor an interaction.

Finally, we have to take into account the meson-meson threshold of Ref.\ \cite{Bali:2005fu} corresponding to two times the static-light meson mass, $E_\text{threshold} = 2 m_{B^{(*)}}$. In the case of two channels we identified $E_\text{threshold}$ with $10.627 \ \text{GeV}$, which is the physical $\bar{B}^{(*)} B^{(*)}$ threshold. However, now using ${m_l}_\text{Ref.\ \cite{Bali:2005fu}} = 22.5 \times m_l$ and performing a linear interpolation between the spin averaged masses of the $B^{(*)}$ meson and the $B_s^{(*)}$ meson, we find $E_\text{threshold} = 10.790 \text{GeV}$.

Thus, the Schr\"odinger equation for the partial wave with $\widetilde{J} = 0$ in the case of three coupled channels is
\begin{widetext}
\begin{eqnarray}
\nonumber & &
\left(-\frac{1}{2} \left(\begin{array}{ccc} 1/\mu_Q & 0 & 0 \\ 0 & 1/\mu_M & 0 \\ 0 & 0 & 1/\mu_{M_s} \end{array}\right) \partial_r^2
+ \frac{1}{2 r^2}
 \left(\begin{array}{ccc} 0 & 0 & 0 \\ 0 & 2/\mu_M & 0 \\ 0 & 0 & 2/\mu_{M_s} \end{array}\right)
+ \left(\begin{array}{ccc} V_{\bar{Q} Q}(r)& V_\textrm{mix}(r) & V_\textrm{mix}(r) / \sqrt{2} \\
V_\textrm{mix}(r) & 0 & 0  \\
V_\textrm{mix}(r) / \sqrt{2} & 0 & 0 \end{array}\right)\right. + \\
 & & \hspace{1.4cm} \left. + \left(\begin{array}{ccc} E_\text{threshold}& 0 & 0 \\
0 & 2 m_M & 0  \\
0 & 0 & 2 m_{M_s} \end{array}\right) - E\right) \left(\begin{array}{c} u(r) \\ \chi_{\bar M M}(r) \\ {\chi}_{\bar M_s M_s}(r) \end{array}\right) =
-\left(\begin{array}{c} V_\textrm{mix}(r)
 \\ 0 \\ 0 \end{array}\right) 
\Big(\alpha r j_1(k r) + \alpha_s r j_1(k_s r) / \sqrt{2}\Big) .
\label{EQN001}
\end{eqnarray}
\end{widetext}
The incident wave can be any linear superposition of a $\bar{B}^{(*)} B^{(*)}$ wave and a $\bar{B}_s^{(*)} B_s^{(*)}$ wave, where $\alpha$ and $\alpha_s$ denote the respective coefficients. For example, a pure $\bar{B}^{(*)} B^{(*)}$ wave translates into $(\alpha,\alpha_s) = (1,0)$ and a pure $\bar{B}_s^{(*)} B_s^{(*)}$ wave into $(\alpha,\alpha_s) = (0,1)$. The momenta of these waves, $k$ and $k_s$, are related to $E$ via 
\begin{eqnarray}
E = 2 m_M + \frac{k^2}{2 \mu_M} \quad , \quad E = 2 m_{M_s} + \frac{k_s^2}{2 \mu_{M_s}} . 
\end{eqnarray}
The corresponding boundary conditions of the wave functions are the following:
\begin{itemize}
\item In both cases (i.e.\ $(\alpha,\alpha_s) = (1,0)$ and $(\alpha,\alpha_s) = (0,1)$):
\begin{eqnarray}
 & & u(r) \propto r \quad \text{for } r \rightarrow 0 \\
 & & u(r) = 0 \quad \text{for } r \rightarrow \infty \\
 & & \chi_{\bar M M}(r) \propto r^2 \ , \ \chi_{\bar M_s M_s}(r) \propto r^2 \quad \text{for } r \rightarrow 0 .
\end{eqnarray}

\item Incident $\bar{B}^{(*)} B^{(*)}$ wave (i.e.\ $(\alpha,\alpha_s) = (1,0)$):
\begin{eqnarray}
\nonumber & & \chi_{\bar M M}(r) = i t_{\bar M M ; \bar M M} r h_1^{(1)}(k r) \ , \\
 & & \chi_{\bar M_s M_s}(r) = i t_{\bar M M ; \bar M_s M_s} r h_1^{(1)}(k_s r) \quad \text{for } r \rightarrow \infty .
\end{eqnarray}

\item Incident $\bar{B}_s^{(*)} B_s^{(*)}$ wave (i.e.\ $(\alpha,\alpha_s) = (0,1)$):
\begin{eqnarray}
\nonumber & & \chi_{\bar M M}(r) = i t_{\bar M_s M_s ; \bar M M} r h_1^{(1)}(k r) \ , \\
 & & \chi_{\bar M_s M_s}(r) = i t_{\bar M_s M_s ; \bar M_s M_s} r h_1^{(1)}(k_s r) \quad \text{for } r \rightarrow \infty .
\end{eqnarray}
\end{itemize}
This defines the $2 \times 2$ matrices $\mbox{S}$ and $\mbox{T}$,
\begin{eqnarray}
\mbox{S} = 1 + 2 i \mbox{T} \quad , \quad \mbox{T} = \left(\begin{array}{cc}
t_{\bar M M ; \bar M M} & t_{\bar M_s M_s ; \bar M M} \\
t_{\bar M M ; \bar M_s M_s} & t_{\bar M_s M_s ; \bar M_s M_s}
\end{array}\right) .
\end{eqnarray}
To determine masses and decay widths of bound states and resonances, we need to find the poles of the $\mbox{S}$ matrix or, equivalently, of the $\mbox{T}$ matrix. We use similar techniques as in our previous work \cite{Bicudo:2019ymo}, but this time we apply the pole search to the determinant of the $\mbox{T}$ matrix.

\begin{figure}[t!]
\includegraphics[width=0.95\columnwidth]{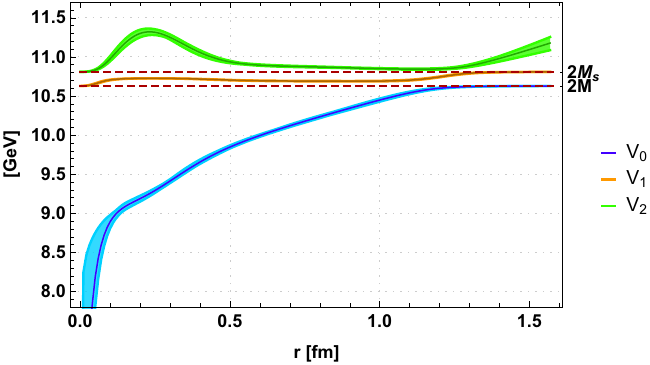}
\caption{\label{fig:potentials3x3}(Color online.) Elements of the diagonalized $3 \times 3$ potential matrix used in the coupled channel Schr\"odinger equation (\ref{EQN001}).}
\end{figure}

It is an interesting consistency check to compare our $3 \times 3$ potential matrix to a recent lattice QCD computation of string breaking with dynamical $u$, $d$ and $s$ quarks \cite{Bulava:2019iut}. For a meaningful comparison we need diagonalize our $3\times 3$ potential matrix. The resulting diagonal elements, which are shown as functions of $r$ in Fig.\ \ref{fig:potentials3x3}, should correspond to the three lowest energy levels of a system with a static quark-antiquark pair and dynamical $u$, $d$ and $s$ quarks.
As expected, they are similar to those plotted in Fig.\ 1 of Ref.\ \cite{Bulava:2019iut}. Note that there is a certain discrepancy in the second excitation at small separations. The bump we obtain and which is not present in Fig.\ 1 of Ref.\ \cite{Bulava:2019iut}, could have different reasons. It might be a consequence of different light quark masses or of the dynamical strange quark used in Ref.\ \cite{Bulava:2019iut} compared to the computation from Ref.\ \cite{Bali:2005fu} or also of imperfect operator optimization. It could also be that our assumptions to set up the $3 \times 3$ potential matrix in Eq.\ (\ref{EQN001}) from the 2-flavor lattice QCD results from Ref.\ \cite{Bali:2005fu} are only partly fulfilled. As we discuss below in our conclusions, we plan to carry out dedicated lattice QCD computations of the relevant potentials in the near future, where we can possibly clarify this tension. For our current work we use the lattice QCD results of Ref.\ \cite{Bali:2005fu}, because numerical values are provided for all required quantities (see Table~I in Ref.\ \cite{Bali:2005fu}). In Ref.\ \cite{Bulava:2019iut}, even though more recent, certain quantities important for our formalism, e.g.\ the mixing angle as a function of $r$, seem not to have been computed.


\section{Quarkonium and meson-meson content of $I = 0$ bottomonium -- two coupled channels\label{sec:content}}

We continue or investigation of bottomonium bound states and resonances with isospin $I = 0$ by studying their structure and quark content. In particular we explore, whether the bound states and resonances close to the $\bar{B}^{(*)} B^{(*)}$ threshold, i.e.\ states with $n = 4, 5, 6$ in Table~\ref{tab:polesemergent}, which could correspond to the experimentally observed $\Upsilon(4S)$, $\Upsilon(10753)$ and $\Upsilon(10860)$, are conventional $\bar{Q} Q$ quarkonia, or whether there is a sizable $\bar{Q} Q \bar{q} q$ four-quark component.
For clarity, we first consider the case of two coupled channels, where it is easier to define the concepts of our study. Then, in Section~\ref{sec:content3}, we will move on to the case of three coupled channels, which is physically more realistic.

We inspect in detail the percentages of quarkonium and of a meson-meson pair present in each of the bound states and resonances. To this end we compute
\begin{eqnarray}
\% \bar Q Q = \frac{Q}{Q + M} \quad , \quad \% \bar M M = \frac{M}{Q + M}
\end{eqnarray}
with
\begin{eqnarray}
\nonumber & & Q = \int_0^{R_\textrm{max}} dr \, \big|u(r)\big|^2 \quad , \quad M = \int_0^{R_\textrm{max}} dr \, \big|\chi_{\bar M M}(r)\big|^2 . \\
 & &
\end{eqnarray}
$u(r)$ and $\chi_{\bar M M}(r)$ are the radial wave functions of the $\bar{Q} Q$ and the $\bar{M} M$ channel, respectively, obtained by solving the coupled channel Schr\"odinger equation (\ref{EQN001_}) with energies $E$ identical to the real parts of the corresponding poles.


\subsubsection{Bound states}

For bound states $E < 2 m_M$ and the corresponding momentum is complex, $k = i \sqrt{|2 \mu_M (E - 2 m_M)|}$. The boundary condition (\ref{EQNbc4}) for $\chi_{\bar M M}(r)$ simplifies to
\begin{eqnarray}
\chi_{\bar M M}(r) = 0 \quad \textrm{for } r \rightarrow \infty .
\end{eqnarray}
Thus, both $Q$ and $M$ are independent of $R_\textrm{max}$, if chosen sufficiently large, i.e.\ $R_\textrm{max} \gtapprox 2.0 \, \textrm{fm}$, because also $u(r) = 0$ for $r \rightarrow \infty$ (see Eq.\ (\ref{EQNbc2})). The same is true for $\% \bar Q Q$ and $\% \bar M M$, which represent the probabilities to either find the system in a quarkonium configuration or in a meson-meson configuration.


\subsubsection{\label{SEC588}Resonances}

For resonances things are more complicated. First, resonances are defined by poles in the complex energy plane with non-vanishing negative imaginary parts of $E$. Evaluating $\% \bar Q Q$ and $\% \bar M M$ at such a complex energy does not seem to be meaningful, because $|u(r)|^2 / r^2$ and $|\chi_{\bar M M}(r)|^2 / r^2$ are only proportional to probability densities, if $E$ is real. Thus we compute $\% \bar Q Q$ and $\% \bar M M$ at the real part of the corresponding pole position, $\textrm{Re}(E)$, which is the resonance mass.

There is, however, another complication, namely that $M$ is not constant but linearly rising for large $R_\textrm{max}$. The reason is that $\chi_{\bar M M}(r)$ represents an emergent wave (see Eq.\ (\ref{EQNbc4})). We found, however, the dependence of $\% \bar Q Q$ and $\% \bar M M$ on $R_\textrm{max}$ to be rather mild, with an uncertainty of only a few percent in the range $1.8 \, \textrm{fm} \le R_\textrm{max} \le 3.0 \, \textrm{fm}$, i.e.\ where the quarkonium component is already negligible, $u(r = R_\textrm{max}) \approx 0$. Thus, we interpret $\% \bar Q Q$ and $\% \bar M M$ as estimates of probabilities to either find the system in a quarkonium configuration or in a meson-meson configuration, as for the bound states discussed before. 


\subsubsection{\label{SEC566}Numerical results}

\begin{figure*}[bht]
\includegraphics[width=0.95\columnwidth]{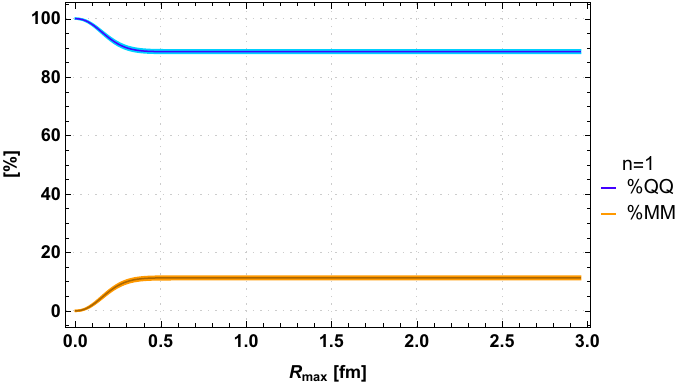}
\includegraphics[width=0.95\columnwidth]{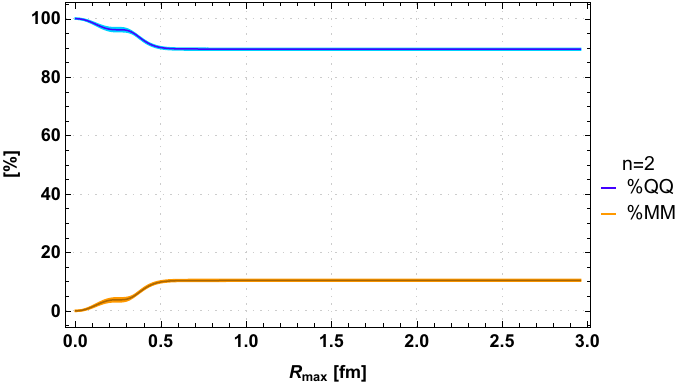}

\vspace{0.2cm}
\includegraphics[width=0.95\columnwidth]{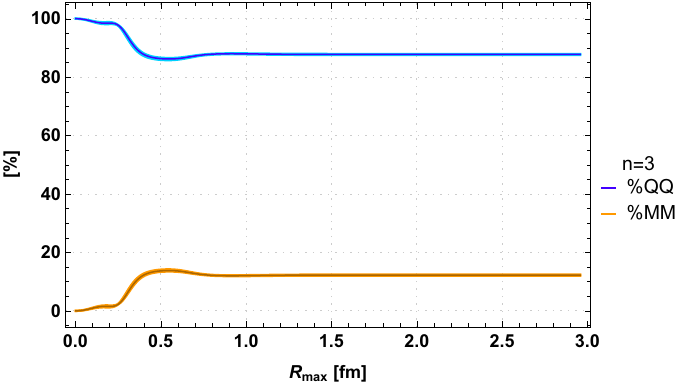}
\includegraphics[width=0.95\columnwidth]{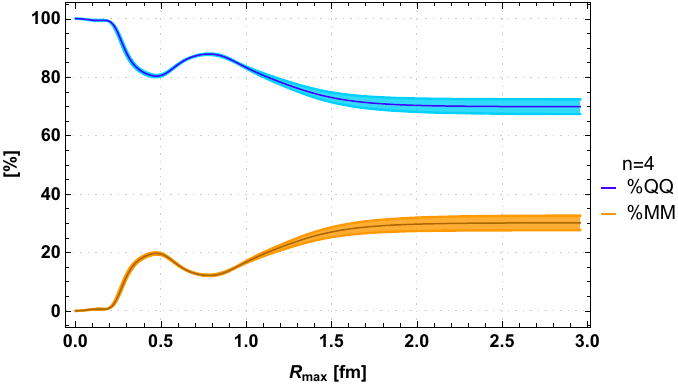}

\vspace{0.2cm}
\includegraphics[width=0.95\columnwidth]{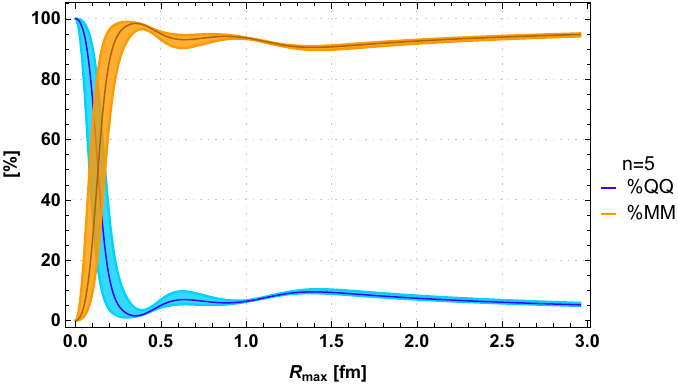}
\includegraphics[width=0.95\columnwidth]{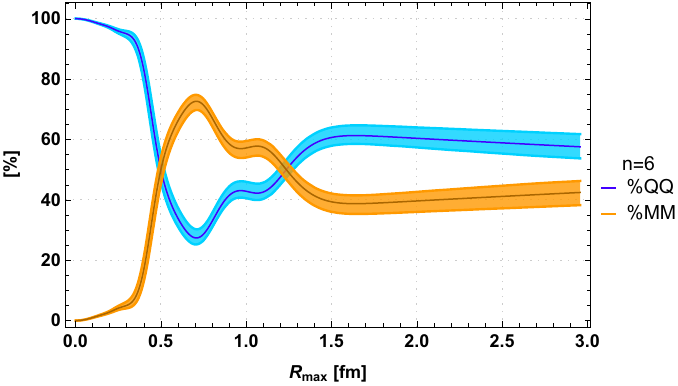}
\caption{\label{fig:percent1000}(Color online.) Percentages of quarkonium $\% \bar Q Q$ and of a meson-meson pair $\% \bar M M$ present in each of the first six bound states and resonances as functions of $R_\textrm{max}$. The error bands represent statistical uncertainties.}
\end{figure*}

We show plots of $\% \bar Q Q$ and $\% \bar M M$ as functions of $R_\textrm{max}$ for the first seven bottomonium bound states and resonances in Fig.\ \ref{fig:percent1000}.

As expected, for the four bound states, $n = 1,\ldots,4$, both $\% \bar Q Q$ and $\% \bar M M$ are constant for large $R_\textrm{max}$. For $\eta_b(1S) \equiv \Upsilon(1S)$ ($n = 1$) this is the case already for $R_\textrm{max} \gtapprox 0.4 \, \textrm{fm}$, while e.g.\ for $\Upsilon(4S)$ ($n = 4$) $R_\textrm{max} \gtapprox 2.0 \, \textrm{fm}$ is needed. This is not surprising and just indicates that wave functions for increasing $n$ are less localized, as usual in quantum mechanics. $\eta_b(1S) \equiv \Upsilon(1S)$, $\Upsilon(2S)$ and $\Upsilon(3S)$ have $\% \bar Q Q \approx 90 \%$, i.e.\ are clearly quarkonium states. $\Upsilon(4S)$, which is close to the $\bar{B}^{(*)} B^{(*)}$ threshold is still quarkonium dominated ($\% \bar Q Q \approx 70 \%$), but already has a sizeable four-quark component ($\% \bar M M \approx 30 \%$).

For the resonances there is a dependence of $\% \bar Q Q$ and $\% \bar M M$ on $R_\textrm{max}$, but it is rather mild with an uncertainty of $2 \%$ or less in the range $1.8 \, \textrm{fm} \le R_\textrm{max} \le 3.0 \, \textrm{fm}$ (see also the discussion in Section~\ref{SEC588}). The wide resonance with $n = 5$ has $\% \bar M M \approx 94 \%$ and, thus, is essentially a meson-meson pair. The resonance with $n = 6$ is a mix of quarkonium and a meson-meson pair with slightly larger $\bar Q Q$ component ($\% \bar Q Q \approx 59 \%$, $\% \bar M M \approx 41 \%$). Resonances with $n \geq 7$ are above the threshold of one heavy-light meson with negative parity and another with positive parity. Since this decay channel is currently neglected, their decay widths are tiny and they are almost stable. Correspondingly, they are strongly quarkonium dominated, i.e.\ $\% \bar Q Q \gg \% \bar M M$. We stress that results for $n \geq 7$ should not be trusted until all relevant decay channels are included.

$\% \bar Q Q$ and $\% \bar M M$ for $R_\textrm{max} = 2.4 \, \textrm{fm}$ are listed in Table~\ref{tab:polesemergent} together with their statistical errors and, for the resonances, also systematic uncertainties. To estimate statistical errors, we utilize the same 1000 sets of parameters as in Ref.\ \cite{Bicudo:2019ymo}, which were generated by resampling the lattice QCD correlators from Ref.\ \cite{Bali:2005fu}. Asymmetric statistical errors are defined via the 16th and 84th percentile of the 1000 samples. We visualize these errors as error bands on $\% \bar Q Q$ and $\% \bar M M$ in Fig.\ \ref{fig:percent1000}. We define the asymmetric systematic uncertainties as $|\% \bar Q Q(R_\textrm{max} = 1.8 \, \textrm{fm}) - \% \bar Q Q(R_\textrm{max} = 2.4 \, \textrm{fm})|$ and $|\% \bar Q Q(R_\textrm{max} = 3.0 \, \textrm{fm}) - \% \bar Q Q(R_\textrm{max} = 2.4 \, \textrm{fm})|$ 	and in the same way for $\% \bar M M$. They are around $2 \%$ for the resonances with $n = 5$ and $n = 6$, respectively, and negligible for all other $n$. The total uncertainties on $\% \bar Q Q$ and $\% \bar M M$ are rather small. Thus, our predictions concerning the structure of the bound states and resonances are quite stable within our framework. The columns ``$\% \bar Q Q$'' and ``$\% \bar M M$'' in Table~\ref{tab:polesemergent} represent the main results for case of two coupled channels, since these numbers reflect the quark composition of the bound states and resonances and clarify, which states are close to ordinary quark model quarkonium, and which states are dynamically generated by a meson-meson decay channel.


\section{Quarkonium and meson-meson content of $I = 0$ bottomonium -- three coupled channels\label{sec:content3}} 

We now consider the case of three coupled channels, a quarkonium, a $\bar{B}^{(*)} B^{(*)}$ and a $\bar{B}_s^{(*)} B_s^{(*)}$ channel.
Working with three channels is technically more elaborate than with two, but formally the extension from the case of two channels is straightforward. To identify the bound states and resonances, we apply our pole searching algorithm \cite{Bicudo:2019ymo} to the determinant of the $\mbox{T}$ matrix. 
In Fig.\ \ref{fig:errorpole} (lower plot) we show the resulting pole positions together with their statistical errors. 

Using the real part of a pole energy, we compute the square of the wave functions of the three channels to determine the relative amount of quarkonium, of a $\bar{B}^{(*)} B^{(*)}$ pair and of a $\bar{B}_s^{(*)} B_s^{(*)}$ pair.
Note that a pole in the $\mbox{T}$ matrix corresponds to one infinite eigenvalue, while the second eigenvalue is finite.
To make a meaningful statement about a bound state or resonance, we thus need to prepare the incident wave in such a way that exclusively the bound state or resonance resonance is generated. This amounts to identifying $(\alpha , \alpha_s)$ appearing on the right hand side of the coupled channel Schr\"odinger equation (\ref{EQN001}) with that eigenvector of $\mbox{T}$ corresponding to the infinite eigenvalue.

\begin{figure*}[htb]
\includegraphics[width=0.95\columnwidth]{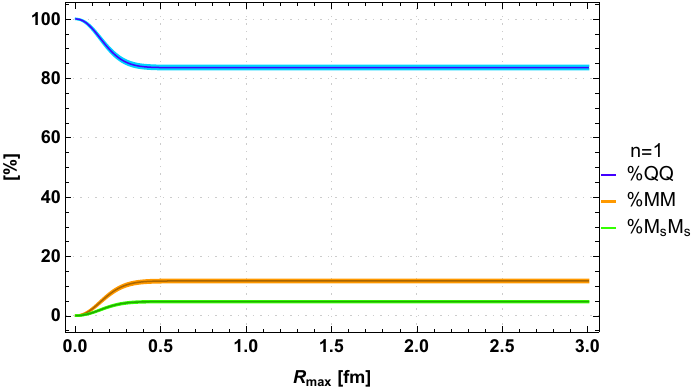}
\includegraphics[width=0.95\columnwidth]{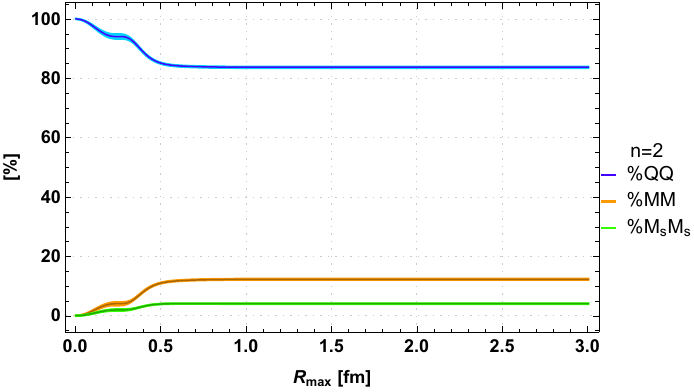}

\vspace{0.2cm}
\includegraphics[width=0.95\columnwidth]{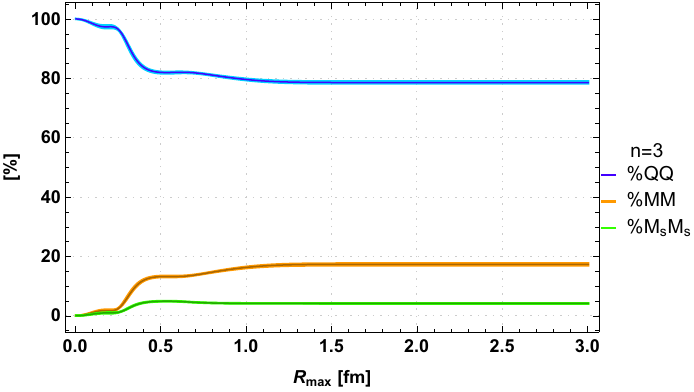}
\includegraphics[width=0.95\columnwidth]{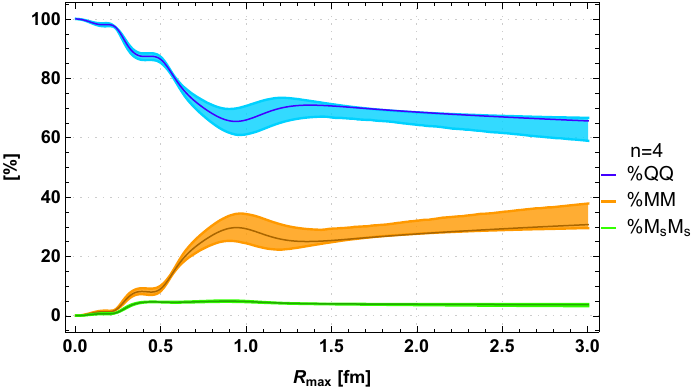}

\vspace{0.2cm}
\includegraphics[width=0.95\columnwidth]{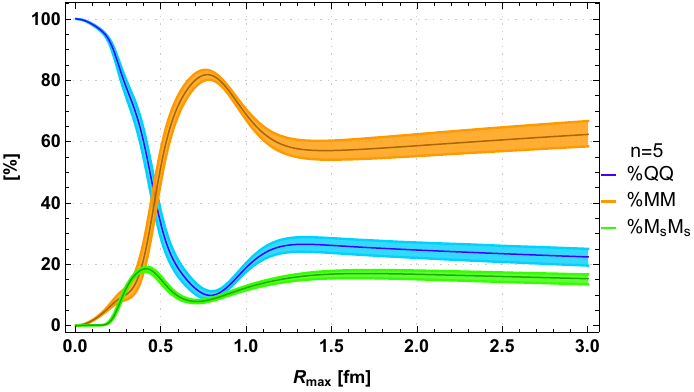}
\includegraphics[width=0.95\columnwidth]{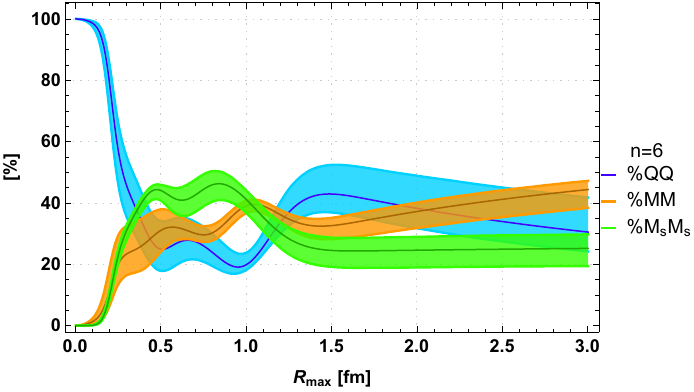}
\caption{(Color online.) Percentages of quarkonium $\% \bar Q Q$ and of meson-meson pairs $\% \bar M M$ and $\% \bar M_s M_s$ present in each of the first six bound states and resonances as functions of $R_\textrm{max}$. The error bands represent statistical uncertainties.
\label{fig:percent3x3}}
\end{figure*}

This time we compute three quantities,
\begin{eqnarray}
\nonumber & & Q = \int_0^{R_\textrm{max}} dr \, \big|u(r)\big|^2 \quad , \quad M = \int_0^{R_\textrm{max}} dr \, \big|\chi_{\bar M M}(r)\big|^2 \quad , \\
 & & \hspace{0.675cm} M_s = \int_0^{R_\textrm{max}} dr \, \big|\chi_{\bar M_s M_s}(r)\big|^2 ,
\end{eqnarray}
from which we calculate the respective percentages of quarkonium and of meson-meson pairs,
\begin{eqnarray}
\nonumber & & \% \bar Q Q = \frac{Q}{Q + M + M_s} \quad , \quad \% \bar M M = \frac{M}{Q + M + M_s} \quad , \\
 & & \hspace{0.675cm} \quad \% \bar M_s M_s = \frac{M_s}{Q + M + M_s} .
\end{eqnarray}
We determine the statistical and systematic errors of the percentages using the same techniques as in Section~\ref{sec:content}. The corresponding results are shown in Fig.\ \ref{fig:percent3x3} as functions of $R_\text{max}$ and also summarized in Table~\ref{tab:polesemergent}.


\subsubsection{Numerical results}

We find that in the three channel case, i.e.\ with a $\bar{B}^{(*)} B^{(*)}$ and a $\bar{B}_s^{(*)} B_s^{(*)}$ channel, the meson-meson percentage increases for the majority of states compared to the two channel case, which has only a $\bar{B}^{(*)} B^{(*)}$ decay channel. 
Nevertheless, the first three states $\Upsilon(1S)$, $\Upsilon(2S)$  and $\Upsilon(3S)$ remain mostly quarkonium states, with $\% \bar Q Q$ around $80 \%$ to $85 \%$. The changes appear to be more pronounced for $n \geq 4$.

The $\Upsilon(4S)$, which is a bound state in the two-channel case, is now a resonance with a decay width more than twice as large as the experimental result. The reason could be that we neglect the heavy quark spins and, thus, the mass of $\Upsilon(4S)$ is not only above the $\bar{B} B$, but also above the $\bar{B}^* B^*$ threshold. Its $\bar{B}^{(*)} B^{(*)}$ content $\% \bar M M \approx 67 \%$ is, however, quite similar to the corresponding percentage obtained in the two-channel case.

In what concerns the new state $\Upsilon(10753)$, the inclusion of the $\bar{B}_s^{(*)} B_s^{(*)}$ channel decreases its decay width from a value much larger than the experimental result to a value consistent with experiment. It remains predominantly a $\bar{B}^{(*)} B^{(*)}$ pair (around $60 \%$), but the quarkonium component increases (to around $24 \%$) and there is now also a non-vanishing $\bar{B}_s^{(*)} B_s^{(*)}$ component (around $16 \%$).

For the $\Upsilon(10860)$, sometimes denominated $\Upsilon(5S)$, the ratio of quarkonium to meson-meson changes from around $59\% / 41\%$ to around $35\% / 65\%$. This is not surprising, because in the three channel case the $\Upsilon(10860)$ is not only above the $\bar{B}^{(*)} B^{(*)}$ threshold, but also above the $\bar{B}_s^{(*)} B_s^{(*)}$ threshold, where the latter increases the meson-meson percentage.

On a qualitative level results obtained with two channels and with three channels are similar. The bound states $n = 1, 2, 3$ consist mostly of quarkonium, while the resonances $n = 4, 5, 6$ have significant meson-meson components. It is particularly noteworthy that there is an additional state compared to the spectrum of pure quarkonium excitations, which is dynamically generated by the coupling to meson-meson decay channels. This state ($n = 5$) has a mass and decay width quite similar to that of the resonance $\Upsilon(10753)$ recently reported by Belle.


\section{\label{sec:conclu}Conclusions}

In Ref.\ \cite{Bicudo:2019ymo} we recently developed 
a novel approach to utilize static potentials computed with lattice QCD in the context of string breaking, opening the way for the computation of the spectrum and the composition of resonances with a heavy quark-antiquark pair and possibly also a light quark-antiquark pair. 
We use these potentials, provided in Ref.\ \cite{Bali:2005fu}, in a coupled channel Schr\"odinger equation, which amounts to applying the diabatic extension of the Born-Oppenheimer approximation, and study the scattering problem with the emergent wave method. In Ref.\ \cite{Bicudo:2019ymo} we coupled a $\bar b b$ quarkonium channel and a $\bar{B}^{(*)} B^{(*)}$ meson-meson channel. In this work we also considered a third channel corresponding to $\bar{B}_s^{(*)} B_s^{(*)}$.

Using this framework we explored the nature of the $I=0$ bottomonium $S$ wave bound states and resonances in more detail, including not only their pole positions but also their compositions in terms of a $b \bar b$ quarkonium component and $\bar{B}^{(*)} B^{(*)} $  
and $\bar{B}_s^{(*)} B_s^{(*)}$ meson-meson components.
This first principles based computation is important, because it contributes to the clarification of controversies concerning the states close to the $\bar{B}^{(*)} B^{(*)}$ threshold 
and the $\bar{B}_s^{(*)} B_s^{(*)} $ threshold (which in our approach are just single thresholds, since the lattice QCD static potentials are independent of the heavy quark spins).

The first controversy concerns the resonances $\Upsilon(10860)$ and $\Upsilon(11020)$. Although they can be identified with $\Upsilon(5S)$ and $\Upsilon(6S)$, they could instead also correspond to the $3D$ or $4D$ states. In our computation we find an $S$ wave state ($n = 6$) somewhat higher, but not too far away from the mass of $\Upsilon(10860)$. Thus, it will be very interesting to also study $D$ wave states within our framework, to see whether there is a better match. In what concerns the $\Upsilon(11020)$ we are currently not in a position to make any reliable statement. Its mass is in the region of the $\bar B^{(*)} B_{0,1}^*$ threshold, i.e.\ the sum of the masses of a negative and a positive parity $B$ meson. Since we do not yet have the lattice QCD potentials to include the coupling to such an excited meson-meson system, the validity of our approach above $\approx 11.025 \, \textrm{GeV}$ is questionable. This is also reflected by the unrealistic imaginary part of the pole we obtain for the resonance with $n = 7$ shown in Fig.\ \ref{fig:errorpole}.

Another controversy concerns the purity as quarkonium states of these resonances, $\Upsilon(10860)$ and $\Upsilon(11020)$, and also of $\Upsilon(4S)$, which is identified according to the Review of Particle Physics \cite{Zyla:2020zbs} as a quarkonium state. We find that $\Upsilon(4S)$ is quarkonium dominated ($\% \bar Q Q \approx 67 \%$), but has a sizable meson-meson component ($\% \bar M M + \% \bar M_s M_s \approx 33 \%$). The $\Upsilon(10860)$, however, is mostly a meson state, composed both of $\bar{B}^{(*)} B^{(*)}$ ($\% \bar M M \approx 40 \%$) and of $\bar{B}_s^{(*)} B_s^{(*)}$ ($\% \bar M_s M_s \approx 25 \%$). In contrast to that, $\Upsilon(1S)$, $\Upsilon(2S)$ and $\Upsilon(3S)$ have rather small meson-meson components, of the order of $15 \%$ to $20 \%$.

The most recent controversy concerns the nature of the newly discovered resonance $\Upsilon(10753)$. Model calculations suggest for instance this resonance to be either a tetraquark \cite{Wang:2019veq,Ali:2019okl}, a hybrid meson \cite{TarrusCastella:2019lyq,Chen:2019uzm,Brambilla:2019esw} or the more canonical and so far missing $\Upsilon(3D)$ \cite{Li:2019qsg,Liang:2019geg,Giron:2020qpb}. With our lattice QCD based approach we find a pole corresponding to the mass $10.773 \, \textrm{GeV}$, similar to the Belle measurement of the mass of the $\Upsilon(10753)$ resonance, $(10.753 \pm 0.007) \, \textrm{GeV}$. In Ref.\ \cite{Bicudo:2019ymo} we had already anticipated this pole to be dynamically generated by the $\bar{B}^{(*)} B^{(*)}$ meson-meson channel. Now, within our improved three channel setup, we confirm that this resonance is mostly composed of a pair of mesons, $\% \bar M M \approx 60 \%$ and $\% \bar M_s M_s \approx 16 \%$. While, there is essentially no direct interaction between a pair of mesons, the mixing potential with the quarkonium channel generates an effective potential sufficiently strong to bind the mesons into a resonance. Thus, since it is not a quarkonium state and the heavy quark spin can be $1^{--}$, it can be classified as a $Y$ type crypto-exotic state. Notice that it should also be part of the $\eta_b$ family, since the heavy quark spin can also be $0^{-+}$ and there is degeneracy with respect to the heavy quark spin.

As an outlook, we are on the way to extend our study beyond $S$ wave bottomonium, to $P$ wave, $D$ wave and $F$ wave, which is more cumbersome, since in these cases there are two additional meson-meson channels. We expect then to be able to address the controversy on the existence of $D$ wave resonances in more detail. Moreover, in the long term we plan to compute lattice QCD static potentials ourselves, in order to update our results with more precision and, hopefully, with excited meson-meson channels, 
possibly even with spin dependent potentials \cite{Lepage:1992tx,Bali:2000gf}.
For example, considering also a $\bar B^{(*)}B_{0,1}^*$ channel with threshold at $\approx 11.025 \, \textrm{GeV}$ would enable us, to predict further excited states not yet discovered in experiments.


\acknowledgements

We acknowledge useful discussions with Gunnar Bali, Eric Braaten, Marco Cardoso, Francesco Knechtli, Vanessa Koch, Sasa Prelovsek, George Rupp and Adam Szczepaniak.

P.B.\ and N.C.\ acknowledge the support of CeFEMA under the FCT contract for R\&D Units UID/CTM/04540/2013 and the FCT project grant CERN/FIS-COM/0029/2017. N.C.\ acknowledges the FCT contract SFRH/BPD/109443/2015.
M.W.\ acknowledges support by the Heisenberg Programme of the Deutsche Forschungsgemeinschaft (DFG, German Research Foundation) -- project number 399217702.


\bibliographystyle{apsrev4-1}
\bibliography{literature.bib}


\end{document}